\documentclass[
	aps, pra, superscriptaddress, twocolumn,
	10pt,
	floatfix, 
    nofootinbib,
	tightenlines, longbibliography
]{revtex4-2}

\usepackage{eucal}
\usepackage{times,bbm,amsmath,amssymb}
\usepackage{epsfig,color}
\usepackage{xcolor}
\usepackage{hyperref}
\usepackage{bm}
\hypersetup{
    colorlinks = true,
    allcolors=blue
}
\usepackage{cleveref}

\usepackage{float,siunitx}
\usepackage[caption = false]{subfig}

\usepackage[greek,english]{babel}
\usepackage{thumbpdf,enumerate}
\usepackage{booktabs}
\usepackage{sidecap}
\usepackage[scaled=.8]{couriers}    
\usepackage{pstricks}
\usepackage{multirow}
\usepackage{placeins}
 \usepackage{relsize}  
\usepackage{pst-grad,bm}
\usepackage{epigraph}
\usepackage{gensymb}
\usepackage{longtable}
\usepackage{soul}
\usepackage{ulem} 
\usepackage{layouts}
\usepackage{acronym}
\usepackage{physics}
\usepackage{easyReview}

\newacro{ML}{Machine Learning}

\usepackage{amsmath,amssymb,amsfonts}
\usepackage{textcomp}
\usepackage{physics}

\usepackage[ruled,linesnumbered]{algorithm2e}

\usepackage{verbatim}
\usepackage{tabularx}

\begin{document}

\title{Witnesses of coherence and dimension from multiphoton indistinguishability tests} 
 
\author{Taira Giordani}
\affiliation{Dipartimento di Fisica, Sapienza Universit\`{a} di Roma, Piazzale Aldo Moro 5, I-00185 Roma, Italy}

\author{Chiara Esposito}
\affiliation{Dipartimento di Fisica, Sapienza Universit\`{a} di Roma, Piazzale Aldo Moro 5, I-00185 Roma, Italy}

\author{Francesco Hoch}
\affiliation{Dipartimento di Fisica, Sapienza Universit\`{a} di Roma, Piazzale Aldo Moro 5, I-00185 Roma, Italy}
\affiliation{Scuola Normale Superiore, Piazza dei Cavalieri, 7 - 56126 Pisa, Italy}

\author{Gonzalo Carvacho} 
\affiliation{Dipartimento di Fisica, Sapienza Universit\`{a} di Roma, Piazzale Aldo Moro 5, I-00185 Roma, Italy}

\author{Daniel J. Brod}
\affiliation{Instituto de F\'{i}sica, Universidade Federal Fluminense, Av. Gal. Milton Tavares de Souza s/n, Niter\'{o}i, RJ, 24210-340, Brazil}

\author{Ernesto F. Galv{\~a}o}
\affiliation{Instituto de F\'{i}sica, Universidade Federal Fluminense, Av. Gal. Milton Tavares de Souza s/n, Niter\'{o}i, RJ, 24210-340, Brazil}
\affiliation{International Iberian Nanotechnology Laboratory (INL), Av. Mestre Jos\'{e} Veiga, 4715-330 Braga, Portugal}

\author{Nicol\`o Spagnolo}
\affiliation{Dipartimento di Fisica, Sapienza Universit\`{a} di Roma, Piazzale Aldo Moro 5, I-00185 Roma, Italy}

\author{Fabio Sciarrino}
\email{fabio.sciarrino@uniroma1.it}
\affiliation{Dipartimento di Fisica, Sapienza Universit\`{a} di Roma, Piazzale Aldo Moro 5, I-00185 Roma, Italy}
\affiliation{Consiglio Nazionale delle Ricerche, Istituto dei Sistemi Complessi (CNR-ISC), Via dei Taurini 19, 00185 Roma, Italy}

\begin{abstract}
%Quantum coherence has been subject for studies on the foundations of quantum mechanics with applications ranging from metrology to cryptography and recently it has proven to be a valuable resource for computation. Finding reliable and effective methods for assessing its presence is then highly desirable. Coherence witnesses are methods consisting of a set of observables whose expected value can guarantee if a state is not diagonal in a known, given basis.
Quantum coherence marks a deviation from classical physics, and has been studied as a resource for metrology and quantum computation.  Finding reliable and effective methods for assessing its presence is then highly desirable. Coherence witnesses rely on measuring observables whose outcomes can guarantee that a state is not diagonal in a known reference basis. Here we experimentally measure a novel type of coherence witness that uses pairwise state comparisons to identify superpositions in a basis-independent way. Our experiment uses a single interferometric set-up to simultaneously measure the three pairwise overlaps among three single-photon states via Hong-Ou-Mandel tests. Besides coherence witnesses, we show the measurements also serve as a Hilbert-space dimension witness. Our results attest to the effectiveness of pooling many two-state comparison tests to ascertain various relational properties of a set of quantum states. %Our findings confirm the effectiveness of this novel family of coherence and dimension witnesses for capturing the quantum properties of high-dimensional systems.
\end{abstract}

\maketitle 

\section{Introduction}
%The existence of superposition states in quantum theory is the key feature that distinguishes it from classical physics.
Quantum coherence is a resource for quantum metrological advantage and quantum computational speed-up, and is central to phenomena such as superfluidity, superconductivity and Bose-Einstein condensation \cite{Streltsov17_rev_coherence, Hillery16}. Its importance has inspired resource-theoretical approaches for its identification and quantification, with the introduction of the idea of coherence witnesses \cite{Baumgratz14_quant_cohe, Napoli16,Piani16}. These are observables whose expectation value provides evidence of superposition, or more precisely, that a quantum state cannot be diagonal in a given reference basis.

\begin{figure*}[t]
    \centering
    \includegraphics[width=\textwidth]{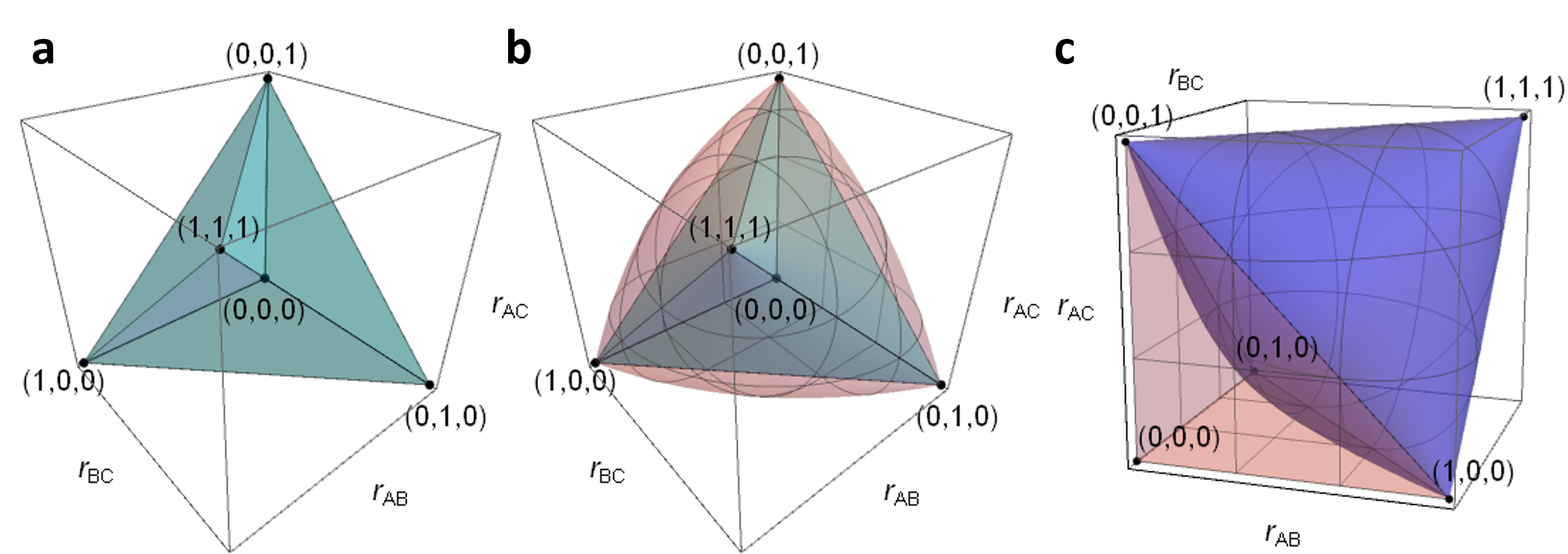}
    \caption{\textbf{Polyhedron $C$ of overlaps compatible with diagonal states, and quantum violations.} \textbf{a)} 5-vertex polyhedron $C$ in the space of 3 overlaps $\vec{r}=(r_{AB}, r_{BC}, r_{AC})$, corresponding to states which are diagonal in a single basis. The convex volumes outside of $C$ correspond to coherent, non-diagonal states. \textbf{b)} The convex body of overlaps corresponding to quantum states (pink) includes the classical polyhedron $C$ (cyan).  \textbf{c)} The convex body $Q_b$ (blue) corresponds to overlaps obtainable from three states spanning a 2-dimensional Hilbert space, and is a strict subset of $Q$. The part of $Q$ outside of $Q_b$ (pink) corresponds to overlaps among states spanning a Hilbert space of dimension $d = 3$.}
    \label{fig:poly}
\end{figure*}

In quantum optics, it is possible to create superposition states of the various degrees of freedom of photons \cite{Flamini_rev}: polarization, path \cite{ Wang_qudits}, transverse mode structure \cite{mair2001entanglement,padgett2004light,nagali2010experimental,sit2017high, Giordani_19qudits}, time-of-arrival \cite{simon2005creating,ali2007large,islam2017provably,de2004long}, frequency \cite{reimer2016generation}, etc. This flexibility, associated with their fast propagation and resistance to decoherence, has resulted in the use of photons for tests of the foundations of quantum theory \cite{de2004long,dada2011experimental, Wang18},  demonstrations of cryptographic key distribution \cite{wang2013direct,islam2017provably, sit2017high, liao2017satellite, liao2017long} and other quantum communication protocols \cite{Cozzolino_rev}. Furthermore, photons represent a significant platform for quantum sensing \cite{pirandola2018metrology} and quantum imaging \cite{genovese2016imaging}. Photons are the perfect information carriers to connect different nodes of a distributed quantum computer, and may even yield scalable, fully photonic quantum computers \cite{qcphoton}.

For many applications, it is important to have photons which are as indistinguishable as possible. The degree of indistinguishability of two photons can be measured by one of the simplest linear-optical devices: a balanced beam-splitter (BS). The famous Hong-Ou-Mandel (HOM) effect \cite{HOM} states that perfectly indistinguishable photons will always leave the balanced BS bunched through the same output port. Various multimode/multiphoton generalizations of the HOM effect have been proposed \cite{Tichy14,PhysRevX.5.041015,Crespi15,Crespi16,Carolan15,Menssen15,Viggianiello18,Giordani18}. Interestingly, an intermediate model of quantum computation was proposed that only uses the interference of multiple single-photons in a multimode, linear interferometer \cite{brod2019review,lund2017review}. 
If we accept some reasonable computational complexity assumptions, these so-called Boson Sampling devices were shown to be hard to simulate by classical computers \cite{AA}, representing a possible approach to demonstrate quantum computational advantage \cite{Harrow2017, Arute2019}.
Recently a theoretical framework was established to use information provided by multiple HOM tests to ascertain the degree of multiphoton indistinguishability \cite{Brod19,Brod19overlaps, Giordani19_4photon}.
HOM experiments are very versatile state-comparison tools, as they provide information on a relational quantity (the overlap), independently on how it is physically encoded in different photonic degrees of freedom \cite{Santori2002,PhysRevLett.106.180501}. HOM tests are a physical, photonic implementation of the abstract quantum circuit known as a SWAP test \cite{chuang2010, Garcia-Escartin,Hendrych2003,Tiedau_2018}. This approach provides a direct advantage for such implementation. Indeed, HOM tests permit to perform SWAP tests with no need for photon-photon interactions, thus not requiring challenging optical nonlinearities.

In this work we extend the cornerstone role of HOM tests by showing how we can pool the information from multiple such tests to establish important relational properties of a set of states. We use a single interferometric set-up to measure the three two-photon overlaps of three single-photon states. We then use the theoretical results of \cite{Brod19overlaps} 
to formulate coherence and dimension witnesses for the case of three states produced by a physical single-photon source. We show that for especially prepared polarization states, our pooled HOM measurements can be used to witness quantum coherence in a basis-independent way, in contrast with basis-dependent witnesses proposed previously,
e.g. in \cite{Baumgratz14_quant_cohe, Napoli16,Piani16, Mogilevtsev_2017}. 
For this, we tuned the two-photon indistinguishabilities to  suitable, intermediate values, which are provably unreachable by coherence-free, diagonal states in any single basis. We also prepare time-of-arrival wavepackets and measure overlap values that work as a dimension witness \cite{Brod19overlaps} that guarantees that the states necessarily span a 3-dimensional Hilbert space. Such dimension witnesses are important in quantum cryptography, where the accidental use of unwanted degrees of freedom represents a security threat \cite{Acin06}.

%Our results highlight how much we can learn from pooling the results of a set of measurements revealing only relational information (the overlaps) among a set of states. Besides being relational, the gathered information is physical, in the sense of revealing two-photon similarity independently of the dimension or type of degree of freedom used to encode information.

\section{Theory of coherence and dimension witnesses from overlaps}
\begin{figure*}[t]
    \centering
    \includegraphics[width=\textwidth]{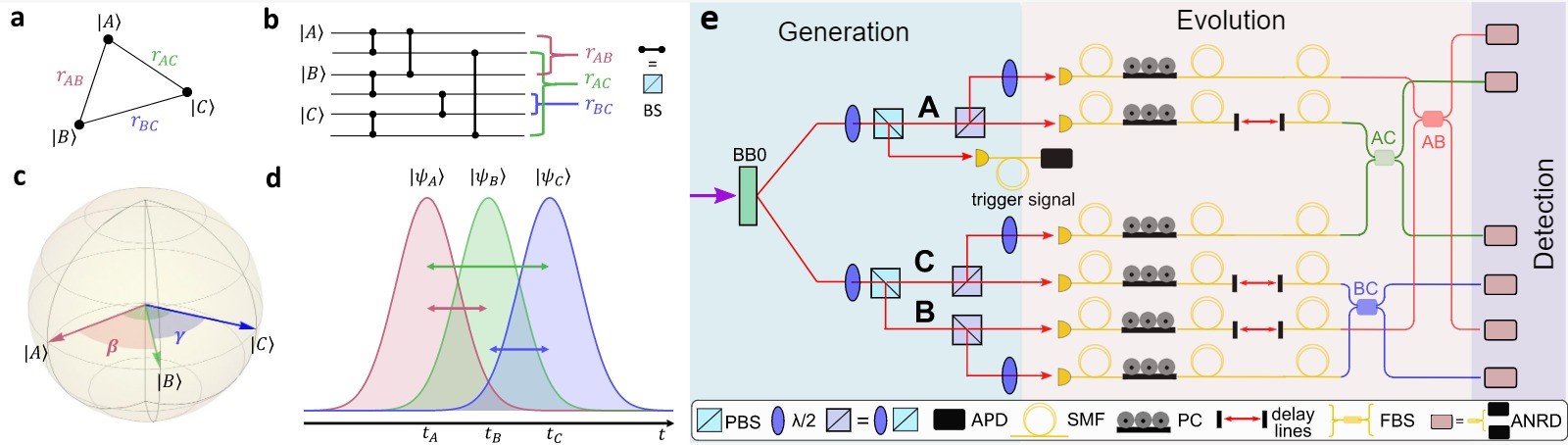}
    \caption{\textbf{Conceptual scheme for measuring coherence and dimension witnesses}. \textbf{a)} Triangle graph $C_3$ represents states (vertices) and overlap measurements (edges). Overlaps $r_{ij}=\textrm{tr}(\rho_i \rho_j)$ are measured via pairwise HOM tests. \textbf{b)}
    All overlaps in $C_3$ can be simultaneously measured using this six-mode linear-optical network. Each overlap is estimated via the probability of bunching at the indicated output modes. \textbf{c)} Bloch-sphere representation of the one-qubit states that maximally violate inequalities (\ref{eq:rineq1})-(\ref{eq:rineq3}). Experimentally, qubits are encoded in the polarization degree of freedom of single-photon states. \textbf{d)} Pictorial model of the time-domain degree of freedom of wavepackets associated with the three photons. Tuning of the relative delays allows for a violation of the dimension witness described in the main text. \textbf{e)} Three-photon states are produced via two-pair generation by spontaneous parametric down conversion (SPDC) in a Beta-Barium Borate (BBO) crystal pumped at $392.5$ nm. One photon serves as the trigger signal and the others, labelled $A$, $B$ and $C$, are prepared for the test. The first interferometer layer in Fig.\ref{fig:concept}-b is realized with bulk beam-splitters made by a half-waveplate ($\lambda/2$) and a polarizing beam-splitter (PBS), (see the legend below). Then photons are coupled into single-mode fibres (SMF), while both polarization and temporal relative delays can be manipulated via additional half-wave plates, %$$$\lambda/2$s
     polarization controllers (PC) and delay lines. In the end, three in-fibre beam-splitters (FBS) connect photon pairs $AB$, $AC$ and $BC$ so as to estimate the overlaps. Each output of FBS is connected to an approximate number-resolving detector (ANRD), consisting of a further FBS and two avalanche-photo-diode detectors (APD).}
    \label{fig:concept}
\end{figure*}
\begin{figure*}[t]
    \centering
    \includegraphics[width=0.95\textwidth]{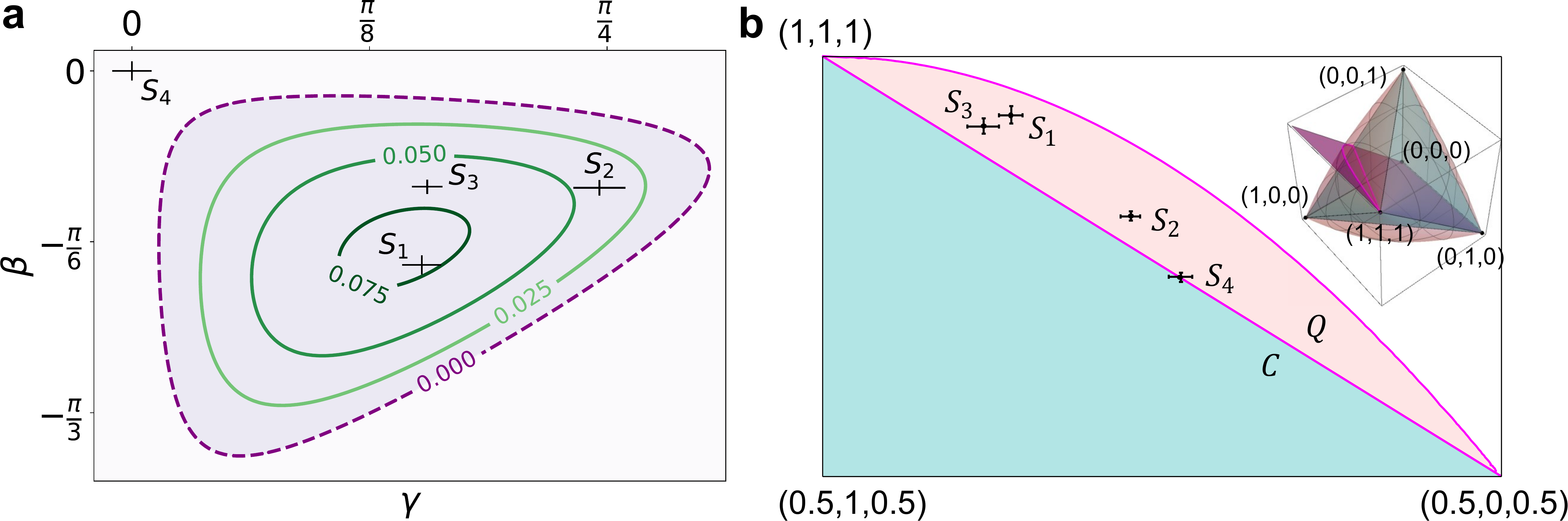}
    \caption{\textbf{Coherence witness of 3 photon polarization states.} \textbf{a)} Amount of violation $W_c(\beta, \gamma)$ of  inequality (\ref{eq:rineq1}), expressed as the Euclidean distance from the closest face of polyhedron $C$, as a function of the angles $\beta$ ($\gamma$) between states $A$-$B$ ($B$-$C$) along a great circle of the Bloch sphere. The purple area corresponds to outcomes preparations for which we can witness coherence ($W_c>0$).  The surface was calculated numerically from the shapes and visibility of the three Hong-Ou-Mandel dips measured during the preparation of the setup. 
    Points $\{S_1,S_2,S_3,S_4\}$ result from suitably prepared polarization states. \textbf{b)} Projection of the four experimental points on the plane perpendicular to the face of $C$ that saturates inequality (\ref{eq:rineq1}) (see inset). The maximum Euclidean distance $W_c$ to $C$ is reached by $S_1$. All the error bars have been retrieved from propagating the Poissonian uncertainties associated to single-photon counts.}
    \label{fig:res}
\end{figure*}
If we restrict quantum mechanics to states and measurements diagonal in a single, reference basis, we recover classical theory. Finding ways to ascertain the presence of superpositions is hence a fundamental problem. Coherence witnesses \cite{Baumgratz14_quant_cohe, Napoli16,Piani16} are observables which can attest that a state is not diagonal in a given, known reference basis. A novel approach to witness superpositions was introduced in Ref.\ \cite{Brod19overlaps}. This method is able to assess coherence without any assumption on the reference basis. 
%in a basis-independent way. 
More precisely, \cite{Brod19overlaps} shows how pairwise overlaps among a set of $n$ states can be used to guarantee that those states cannot be diagonal in \textit{any} single basis. Moreover, the tests are experimentally friendly, as overlaps can be estimated by a simple quantum circuit called the SWAP test \cite{chuang2010, Garcia-Escartin}.

Let us now recall the theory behind the simplest %such
setup to witness quantum coherence. 
%Let us 
Consider three (generally mixed) quantum states $\rho_A$, $\rho_B$ and $\rho_C$, and assume they are coherence-free, i.e.\ they are diagonal in a single basis formed by eigenvectors $\{ \ket{\phi_i}\}$ of a reference observable $\hat{O}$. The pairwise overlap $r_{AB}\equiv \textrm{tr} (\rho_A \rho_B)$ is then just the probability of obtaining the same outcome $v(\hat{O})$ when observable $\hat{O}$ is measured separately on states $\rho_A$ and $\rho_B$
(and similarly for the other two overlaps $r_{AC}$ and $r_{BC}$).
Now consider a single-shot experiment which measures $\hat{O}$ separately on $\rho_A, \rho_B$ and $\rho_C$. Let us describe the states associated with the obtained measurement outcomes using the three overlaps $\vec{r}=(r_{AB}, r_{AC}, r_{BC})$. It is clear that if two of the overlaps are 1, then the third overlap also must equal 1, by the transitivity of equality. So there are 3 possibilities for $\vec{r}=(r_{AB}, r_{AC}, r_{BC})$ which are logically impossible: $\{ (1,1,0), (1,0,1), (0,1,1)\}$. The possible outcomes must then be described by one of the five allowed states $\vec{r} \in \{ (0, 0, 0),(0, 0, 1),(0, 1, 0),(1, 0, 0),(1, 1, 1) \}$.

So far we have only illustrated a single-shot experiment. After performing many measurements of $\hat{O}$ on the three states, we will in general obtain probabilistic results, described by $\vec{r}$ which must lie in the convex hull of the five allowed, deterministic $\vec{r}$ we have listed above. This polyhedron $C$ corresponding to overlaps obtainable from three diagonal, coherence-free states can be seen in Fig. \ref{fig:poly}a. Note that besides the trivial inequalities satisfied by the overlaps ($r_{ij}\ge 0, r_{ij} \le 1$), there are 3 non-trivial new inequalities: 
\begin{eqnarray}
r_{AC} \ge r_{AB}+r_{BC}-1, \label{eq:rineq1}\\
r_{AC} \le r_{AB}-r_{BC}+1, \label{eq:rineq2}\\
r_{AC} \le r_{BC}-r_{AB}+1. \label{eq:rineq3}
\end{eqnarray}
To recapitulate: the three inequalities above are a consequence of simple logic, and refer to the probabilities of obtaining the same outcome when measuring observable $\hat{O}$ independently on pairs of states, all of which assumed to be simultaneously diagonal in the basis of $\hat{O}$. 
The latter assumption does not hold for all quantum states.
Indeed quantum mechanics can violate the above inequalities.
While this is logically impossible by simultaneous, direct measurements of $\hat{O}$ on the three states, as we will soon see it is possible to do an indirect measurement of the three overlaps, with results that may violate inequalities (\ref{eq:rineq1}-\ref{eq:rineq3}). 
In this sense a violation serves as a basis-independent coherence witness of the three states.

In Fig. \ref{fig:poly}b we have a geometrical representation of the polyhedron $C$ of overlap triples arising from  diagonal, coherence-free states, enclosed by the set $Q$ of overlap triples arising from general quantum states.
The set $Q$ is defined by three bounds, which can be written
\begin{equation}
   r_{AC}\ge \left(\sqrt{r_{AB}r_{BC}}-\sqrt{(1-r_{AB})(1-r_{BC})}\right)^2 
   \label{eq:rineq_qb}
\end{equation}
if $r_{AB}+r_{AC}>1$, and $r_{AC}\ge 0$ otherwise, together with two equivalent expressions obtained by permutations of the indices $A,B$, and $C$. The derivation of these inequalities can be found in Ref.\ \cite{Brod19overlaps}.

Points $\vec{r}$ contained in $Q$ but not in $C$ correspond to triplets of states displaying quantum coherence. The points in $Q$ farthest from the boundary of $C$ correspond to a maximal violation of inequalities (\ref{eq:rineq1}-\ref{eq:rineq3}), and one such example can be obtained by three one-qubit states separated by $60^o$ on a great circle of the Bloch sphere (see Fig. \ref{fig:concept}-c).

The measurement of a set of two-state overlaps can also be used as a dimension witness, establishing a lower bound for the dimension of the Hilbert space spanned by a set of states. This is possible because the set of overlaps $\vec{r}$ attainable by three states spanning a 2-dimensional Hilbert space is strictly smaller than the set attainable by states spanning a 3-dimensional Hilbert space. 
If three states span only a two-dimensional Hilbert space, inequality (\ref{eq:rineq_qb}) holds always, rather than only when $r_{AB}+r_{AC}>1$ (and the same holds for the two other inequalities obtained by permutation of indices) \cite{Brod19overlaps}. This defines a new region $Q_b \subset Q$. Intuitively, this happens because the point $\vec{r}=(0,0,0)$ is no longer allowed---three single-qubit states cannot be mutually orthogonal. This can be seen in Fig.\ \ref{fig:poly}-c, where we plot a comparison between $Q$ and $Q_b$. It is clear that $Q_b$ differs from $Q$ in the vicinity of point $(0,0,0)$.

Therefore, observing a set of overlaps corresponding to a point outside of $Q_b$ serves as a witness that the Hilbert space spanned by the three states is 3-dimensional.This type of photonic dimension witnesses may find applications in quantum machine learning - recognizing linear dependence among a set of quantum states allows for a compressed description of the data (see e.g. \cite{LloydMR14,Ghosh2019}).

\section{Experimental setup and results}
In the more abstract formalism of quantum circuits with qubits, the overlap $r_{AB}$ of two given states $\rho_A, \rho_B$ can be measured by projecting onto the symmetric subspace, which can be done by a simple quantum circuit known as a SWAP test \cite{chuang2010}. Specifically for photonics, there is a simple way to implement SWAP tests, namely via two-photon, Hong-Ou-Mandel interferometric experiments \cite{Garcia-Escartin}. In a HOM test, two photons impinge on different input ports of a 50/50 beam-splitter, and the probability that photons come out bunched in a single output port $p_b$ is given by $p_b=(1+r_{AB})/2$ \cite{Garcia-Escartin, Brod19, Giordani19_4photon}.
The needed HOM tests can be simultaneously performed using a single multimode interferometric setup. This is exemplified for the case of simultaneous estimation of three pairwise overlaps of three states in Fig.\ \ref{fig:concept}a-b. This provides the full data required to test the coherence witness using inequalities (\ref{eq:rineq1})-(\ref{eq:rineq3}), and the dimensionality witness described before.

\subsection{Experimental apparatus}
Let us now briefly describe the platform that enables the implementation of coherence and dimension witnesses using photonic states. The quantum states describe three photons generated by a spontaneous parametric down-conversion (SPDC) source, comprising a BBO crystal operating at $785$ nm. We focus on the generation of two photon pairs, with one photon being used as to herald the injection of the other three in the interferometer of Fig. \ref{fig:concept}-b (see App. \ref{app:preliminary}). Measurements record four-fold coincidences of the herald and signal photons. The apparatus is described schematically in Fig. \ref{fig:concept}e. Each beam-splitter in the first interferometer layer is realized with a half wave-plate ($\lambda/2$) followed by a polarizing beam-splitter (PBS). Then we have a further half wave-plate on each beam reflected by the PBS, and polarization controllers (PC) on the single-mode fibre (SMF) to tune pairwise overlaps with polarization (see App. \ref{app:polarization} for further details). After these stages, delay lines allow us to control the temporal overlaps between wave-packets. The final in-fibre beam-splitters (FBSs) implement the three Hong-Ou-Mandel tests needed to calculate the three overlaps. The detection stage employs 12+1=13 single-photon detectors. Each output of FBSs in Fig. \ref{fig:concept}e is split into two further auxiliary modes in order to obtain an approximate photon-number resolving detector. This is required, as the overlap is estimated from the probability of bunching, i.e. the probability that two photons exit the FBS bunched in a single output port.

\subsection{Pairwise overlap model}
Quantum states are commonly encoded in different degrees of freedom of single photons. Here we exploit the polarization for encoding qubits, using temporal relative delays to enlarge the dimension of the Hilbert space describing the systems. Hence, for our aims, we need to provide an explicit expression for the pairwise overlaps in terms of these two degrees of freedom. We model each overlap as a product of three factors: the overlap between the polarization vectors; the temporal overlap of the photons' wave-packets; and then a quantity that accounts for overlaps of the remaining degrees of freedom (e.g. in the frequency domain). For each pair of photons connected by the graph of Fig. \ref{fig:concept}-a, we then have:
\begin{equation}
    r_{ij}=V_{ij}\cdot \lvert\langle e_i|e_j\rangle \lvert^2 \cdot \lvert\langle \psi_i(t_i)|\psi_j(t_j)\rangle\lvert^2
    \label{eq:r_exp}
\end{equation}
where $\ket{e_i}$ stands for the polarization state, $\ket{\psi_i(t_i)}$ describes the photon's wave-function in the time domain, and $V_{ij}$ is the integral corresponding to the overlap over other degrees of freedom.

The maximal violation that witnesses coherence can be obtained by three single-qubit states on a great circle of the Bloch sphere \cite{Brod19} (see Fig.\ \ref{fig:concept}-c). By using the linear polarized states $\{\ket{H},\ket{V}\}$ as a basis, one convenient choice of great circle are states of the form $\ket{e(\theta)} = \cos{\theta}\ket{H}+\sin{\theta}\ket{V}$, with $\theta \in [0, \pi/2)$. The overlap between two such states is given by $\left \lvert\langle e(\theta)\lvert e(\phi)\rangle\right\lvert^2=\cos^2(\theta-\phi)$.
The states $\ket{e(\theta)}$ can be generated from the $\ket{H}:=\ket{e(0)}$ state by the application of a single half-wave plate, whose action results in a rotation of the vectors $\ket{e_i}$ by an angle $2\alpha$, where $\alpha$ is the inclination of the optical axis of the material with respect the $\ket{H}$ direction. 
In our setup (see Fig.\ \ref{fig:concept}-e), the three photons are prepared in the state $\ket{H}$. To achieve that, we begin with a layer of PBS that selects the photons by their linear polarization, and for those photons in the reflected arms (e.g.\ which are in the $\ket{V}$ state) we compensate with a half-wave plate to set them to  $\ket{H}$. The axes of the half-waveplates are then further rotated to generate single-qubit states with the desired polarization overlaps. To ensure such level of control over polarization states, it was necessary to take into account the effect of single-mode fibres, which was done via a preliminary compensation of the rotations induced by the input arms of FBSs, $U_i$ and $U_j$ respectively. This was achieved by imposing $U_i=\mathrm{e}^{\mathrm{i}\phi} U_j$, where $\phi$ is a global phase, thanks to the in-fibre polarization control in the setup.

The temporal contribution to $r_{ij}$ can be expressed as $\lvert\langle\psi_i(t_i)|\psi_j(t_j)\rangle\lvert^2= e^{-\sigma_{ij}(t_i-t_j)^2}$ where $\sigma_{ij}$ are the widths of the mutual overlaps of single-photon spectra, modelled as Gaussian. 
Both $\sigma_{ij}$ and $V_{ij}$ were estimated by preliminary Hong-Ou-Mandel tests between the pairs, as discussed in App. \ref{app:preliminary}. The temporal overlaps between photons are controlled by suitable changes in the delay line positions.
The three independent temporal delays are calibrated to generate logically allowed states.

\subsection{Coherence witness using three polarization states}
\begin{figure}
    \centering
    \includegraphics[width=\columnwidth]{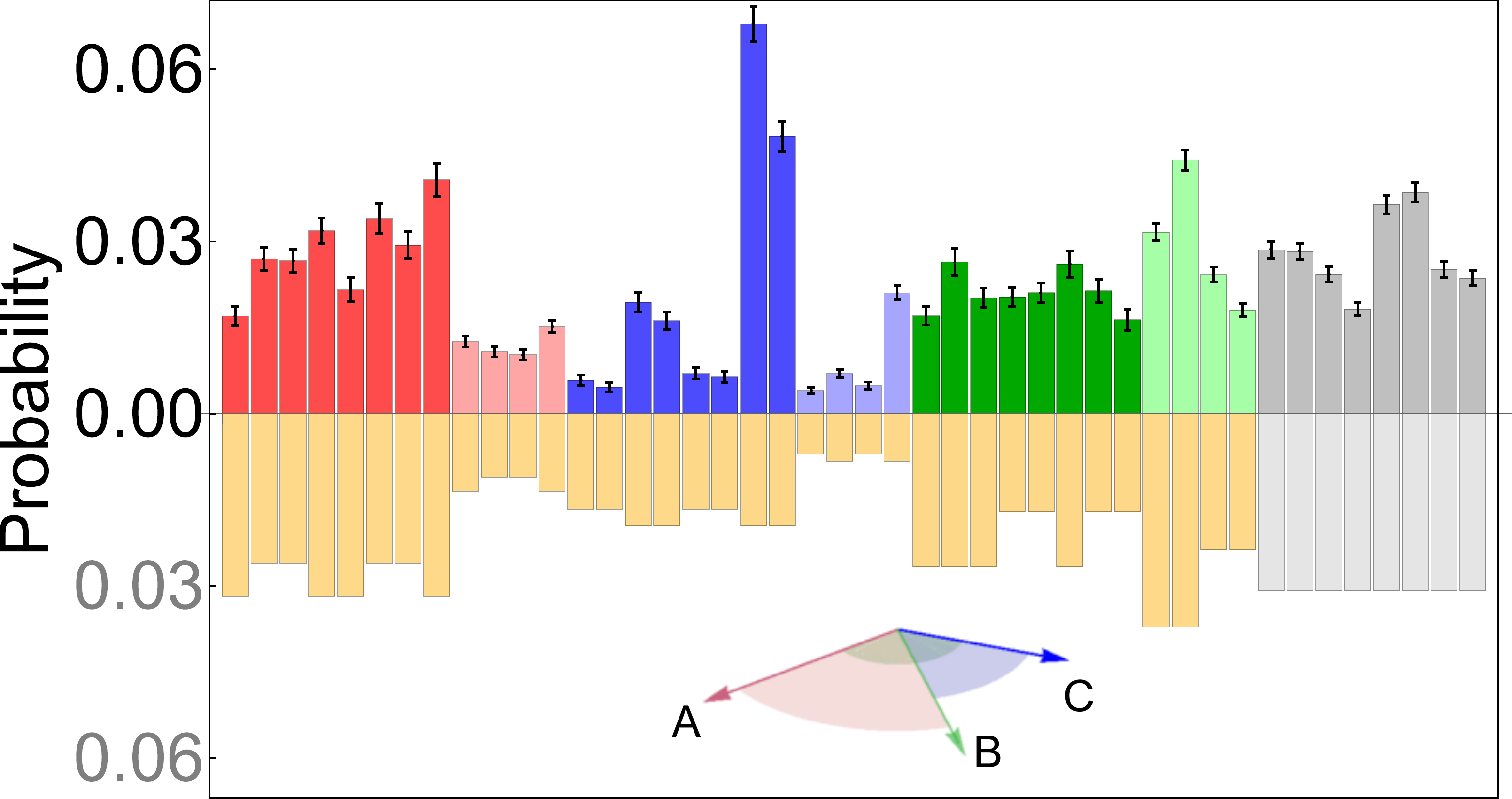}
\caption{\textbf{Output probability distribution for set $S_1$ of polarization states.} 
 Above: experimental data, where brighter colors indicate bunched outcomes. The x-axis reports the three-photon outcomes at the output of the six-port interferometer. Outcomes associated with photon pairs $\{(A,B),\, (B,C),\, (A,C)\}$ are marked respectively with red, blue and green. The grey part of the distribution is insensitive to two-photon interference and so it is irrelevant for the estimation of overlaps. Below: expected distribution according to our model based on preliminary measurements of Hong-Ou-Mandel dips. We consider the total variation distance (TVD) as a quantifier of the agreement with the theory. Error bars are computed propagating, via Monte Carlo methods, the Poissonian uncertainties associated to photon counts. We collected $N\sim 10^4$ four-fold coincidence and we obtained a TVD$=0.138\pm0.005$. In the inset, the three single-photon states on the equator of the Bloch sphere.}
   \label{fig:distr}
\end{figure}
In Fig.\ref{fig:poly}-a we show the polyhedron $C$ of overlap triples $\vec{r}=(r_{AB}, r_{BC}, r_{AC})$ compatible with three states diagonal in a common basis. Our goal is to prepare the system in states which lie outside one of the non-trivial faces of $C$: we pick states that violate the face $r_{AC}=r_{AB}+r_{BC}-1$ of $C$, and define $W_c$ as the Euclidean distance of an overlap triple to that face.

We can now investigate the conditions for which our pairwise overlap model \eqref{eq:r_exp} predicts the possibility of witnessing coherence. To this end we express $W_c$ as a function of the parameters $\{\beta, \gamma\}$ (the angles between state $\lvert  B \rangle$ and states $\{\lvert  A \rangle,\lvert  C \rangle\}$, as in Fig.\ \ref{fig:concept}-c), and whose value also depends on the $V_{ij}$ and the temporal overlaps between wavepackets. If we now assume perfect synchronization of the three wavepackets  ($t_i=t_j$ for all pairs  $\{i,j\}$), we have
 \begin{equation}
 \begin{split}
     W_{c}(\beta, \gamma)=&\frac{1}{\sqrt{3}}\vert 1- V_{AB}\cos^2{\beta} +\\ &-V_{BC}\cos^2{\gamma}+ V_{AC}\cos^2(\beta-\gamma) \vert .
 \end{split}
 \end{equation}
In Fig.\ \ref{fig:res}-a we show the contour lines of values of $W_c(\beta, \gamma)$, highlighting in purple the region that leads to triples outside of $C$, i.e for which $W_c>0$. We estimate a maximum value of $W_c=0.08$ according to our model of the system. Note that overlap triples in the white area cannot guarantee coherence, as they are compatible with coherence-free states $\in C$.

Following this numerical investigation, we prepared four sets of single-qubit polarization states corresponding to points $\{S_1, S_2, S_3, S_4\}$ in Fig.\ \ref{fig:res}-a. Point $S_4$ corresponds to a set of three equal states, i.e.\ $(\beta,\gamma)=(0,0)$, which is in $C$. We then moved on to prepare states for which it is possible to witness coherence: $S_1$ corresponds to the pair $(\beta,\gamma)$ that leads to a maximum violation of the coherence witness, while $S_2$ and $S_3$ correspond to two sets of polarization configurations for which we expected lower witness violations. We report the sets of qubit states first according to the values of $(\beta, \gamma)$ in Fig.\ \ref{fig:res}-a, which were obtained from the experiments by the relations $|\beta|=\arccos{\left(\sqrt{\frac{r_{AB}}{V_{AB}}}\right)}$ and $|\gamma|=\arccos{\left(\sqrt{\frac{r_{BC}}{V_{BC}}}\right)}$. Then we report the points in the three-dimensional space of the overlaps, highlighting the position with respect to polyhedron $C$ and the quantum body $Q$. For easier visualization, we project the points and the two borders of $C$ and $Q$ on the plane perpendicular to the relevant face of $C$ (see Fig.\ref{fig:res}-b).
The highest violation is by point $S_1$, which was tailored to maximize the distance to the closest face of $C$, thus confirming the agreement between our model and the experiment. The distance from $S_1$ to the closest boundary of $C$ is $W_{c_1}=0.081 \pm 0.014$, which is positive by $5.7$ standard deviations. The values for $S_1 \ldots S_4$ are reported in Fig.\ \ref{fig:res}-b and in Table \ref{table:qubit}.
In Fig. 4 we report the three-photon output distribution corresponding to input configuration $S_1$. The distributions of the other polarization states have been included in App. \ref{app:distributions}.
\begingroup
\setlength{\tabcolsep}{3pt} % Default value: 6pt
\renewcommand{\arraystretch}{1.5} 
\begin{table}[h!]
   \centering
    \begin{tabular}{|c c|c |c|c|}
    \hline
    \hline
    & $r_{AB}$ &  $r_{BC}$& $r_{AC}$ &  $W_c$\\
    \hline
    \hline
   $S_1$ & $0.648\pm0.014$ &\,$0.63\pm0.01$ & $0.14\pm0.02$  & $0.081 \pm 0.014$\\
   $S_2$ & $0.830\pm0.008$ &\,$0.624\pm0.012$ & $0.381\pm0.012$  & $0.041 \pm 0.010$\\
 $S_3$ &$0.828\pm0.009$ &\,$0.409\pm0.021$ & $0.167\pm0.019$ &   $0.040 \pm 0.017$\\
$S_4$ & $0.827\pm0.011$ &\,$0.701\pm0.014$ & $0.527\pm0.013$ &  $0.000 \pm 0.010$\\
    \hline
    \hline
    \end{tabular}
     \caption{\textbf{Measured overlaps and distance from coherence-free polyhedron $C$.} Overlaps $\{r_{AB}, r_{BC}, r_{AC}\}$ corresponding to the 4 sets of states generated in the experiment. 
     Column $W_c$ is the Euclidean distance from the closest face of polyhedron $C$, i.e. the amount of violation of inequality (\ref{eq:rineq1}).}
     \label{table:qubit}
\end{table}
\endgroup

\begin{figure}[t]
    \centering
    \includegraphics[width=\columnwidth]{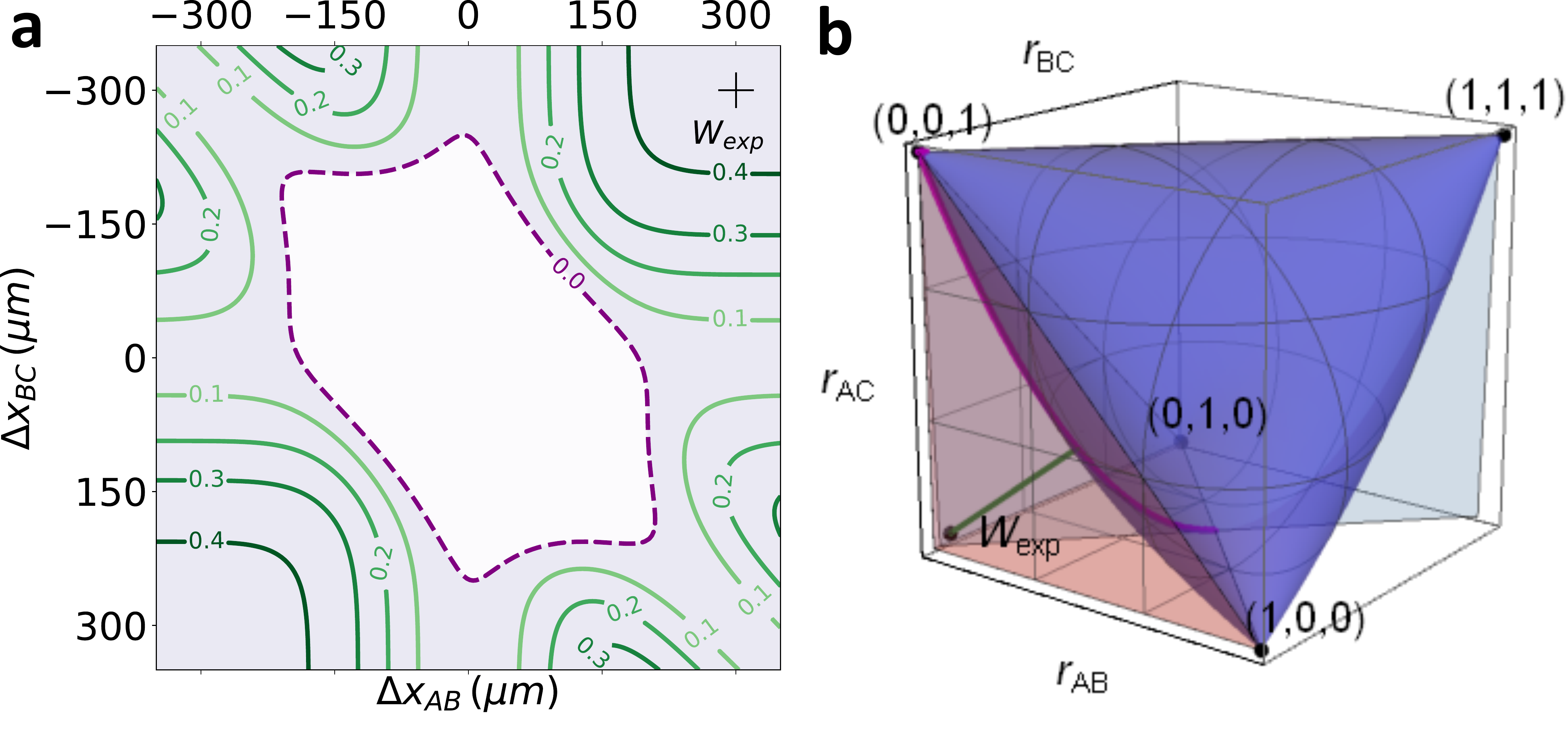}
    \caption{\textbf{Dimension witness $W_d$.} $W_d$ is the minimum Euclidean distance from an overlap triple $(r_{AB},r_{BC}, r_{AC})$ to the body $Q_b$ of triples compatible with states spanning only a 2-dimensional Hilbert space. \textbf{a)} $W_d$ as a function of the two relative time delays between the three photons, in the presence of experimental imperfections. The purple area corresponds to $W_d>0$, i.e. triples $\notin Q_b$. \textbf{b)} Geometrical representation of the set $Q_b$ (blue), and the set of overlap triples only reachable by states spanning a 3-dimensional Hilbert space (magenta). The plane in light blue includes the experimental datum and the green line segment (of length $W_d$) connecting it to the closest point in $Q_b$. 
    }
    \label{fig:dim_witness}
\end{figure}

\subsection{Dimension witness from relative temporal delays}
As discussed previously, our characterization of the regions of allowed overlap triples $(r_{AB},r_{BC},r_{AC})$ can also be used as a dimension witness. To that end we focus on a different region than in the case of the coherence witness. Now we are interested in the difference between the regions accessible by general quantum states ($Q$) and by states spanning a Hilbert space which is at most 2-dimensional ($Q_b$) [cf. Fig.\ref{fig:poly}c]. By observing an overlap triple in $Q$ but not in $Q_b$, we  confirm experimentally that the Hilbert space spanned by our three states is necessarily three-dimensional. As discussed previously, we require a set of states for which inequality (\ref{eq:rineq_qb}) is violated when $r_{AB}+r_{BC} \leq 1$.

A simple choice for this purpose corresponds to three mutually orthogonal states, unreachable if states do not span a 3-dimensional Hilbert space. Even though experimental fluctuations mean that an exactly null overlap will never be measured, our bounds enable us to rule out the possibility of a smaller Hilbert space span in a robust way, despite the fluctuations.

In our experimental set-up, high-dimensional states were created by introducing delays between three single-photon wavepackets with identical polarizations (see Fig.2-d). Our dimension witness $W_d$ is the distance between the resulting overlap triple $(r_{AB},r_{BC},r_{AC})$ and the body $Q_b$, of triples consistent with states that span a 2-dimensional Hilbert space. $W_d>0$ corresponds to points $\notin Q_b$. In Fig.\ \ref{fig:dim_witness}-a we show how $W_d$ depends on the two wave-packet delays. In Fig.\ \ref{fig:dim_witness}-b we see that the resulting triple is clearly outside of $Q_b$. In Table \ref{tab:dim_witness} we report the resulting overlap triple retrieved from the three-photon distribution shown in the App. \ref{app:distributions}, resulting in a distance $W_d=0.380 \pm 0.019$ to $Q_b$. The measured overlap triple is inconsistent with the hypothesis that states span only a 2-dimensional Hilbert space by $\sim 20$ standard deviations.

\begin{table}[t]
    \centering
    \begin{tabular}{|c|c |c |c |}
    \hline
    \hline
    $r_{AB}$ & $r_{BC}$ & $r_{AC}$  & $W_d$ \\
    \hline
    \hline
    $0.019\pm0.021$ & $0.041\pm 0.025$ & $0.032\pm0.02$ &  
    $0.380\pm0.019$\\
    \hline
    \hline
    \end{tabular}
    \caption{\textbf{Measured overlaps incompatible with states spanning a 2-dimensional Hilbert space.}
    We report the overlaps measured for three time-delayed wavepackets. $W_d$ represents the minimum Euclidean distance between the experimental point $x_{\mathrm{exp}}=(r_{AB},r_{BC}, r_{AC})$ and the closest point  of the body $Q_b$ of overlap triples compatible with states spanning a 2-dimensional Hilbert space. 
    }
    \label{tab:dim_witness}
\end{table}

\section{Discussion}

Whether or not a quantum state is a superposition depends, of course, on the reference basis chosen. When we fix a reference basis, coherence witnesses have been proposed as a convenient way to detect superposition, using a simpler set-up than full state tomography. When we have three or more states, however, the very geometry of their pairwise overlaps can be used to guarantee the presence of quantum coherence in a basis-independent way. Here we have experimentally demonstrated this approach to witness quantum coherence via pairwise overlap measurements on a set of three photonic states. It would be interesting to develop further the theory of such coherence witnesses, for example by characterizing how the amount of violation can quantify coherence, and its usefulness in quantum computation. We have also used overlap measurements to demonstrate a Hilbert space dimension witness. Such witnesses can be useful in the analysis of quantum key distribution schemes, where unwanted correlations in a space of higher dimension can result in a security threat.

Our coherence and dimension witnesses rely on pooling information from a set of two-state SWAP tests, which can be conveniently implemented as Hong-Ou-Mandel tests in a linear-optical setup. Our schemes can be extended in a straightforward way to larger number of states. Furthermore, the scheme is not limited to the presented encoding using polarization and time. Thus, it can be also employed in a more general scenario involving different degrees of freedom and variable number of states. This suggests that our results may find applications in near-term linear optical computers using SWAP test-based quantum algorithms for distance calculations and data clustering \cite{Aimeur06, Lloyd14, Wiebe15}.

\begin{acknowledgments}
We acknowledge funding from the European Research Council (ERC) Advanced Grant QU-BOSS (QUantum advantage via non-linear BOSon Sampling, grant agreement no. 884676), the European Research Council (ERC) Advanced Grant CAPABLE (Composite integrated photonic platform by femtosecond laser micro-machining, Grant Agreement No. 742745), from the QuantERA ERA-NET Cofund in Quantum Technologies 2017 project HiPhoP (High dimensional quantum Photonic Platform, project ID 731473), and from the Instituto Nacional de Ci\^{e}ncia e Tecnologia de Informa\c{c}\~{a}o Qu\^{a}ntica (INCT-IQ/CNPq).
\end{acknowledgments}

\appendix

\section{Preliminary characterization of the single-photon source}
\label{app:preliminary}
\begin{figure}[t!]
    \centering
    \includegraphics[width=0.7\columnwidth]{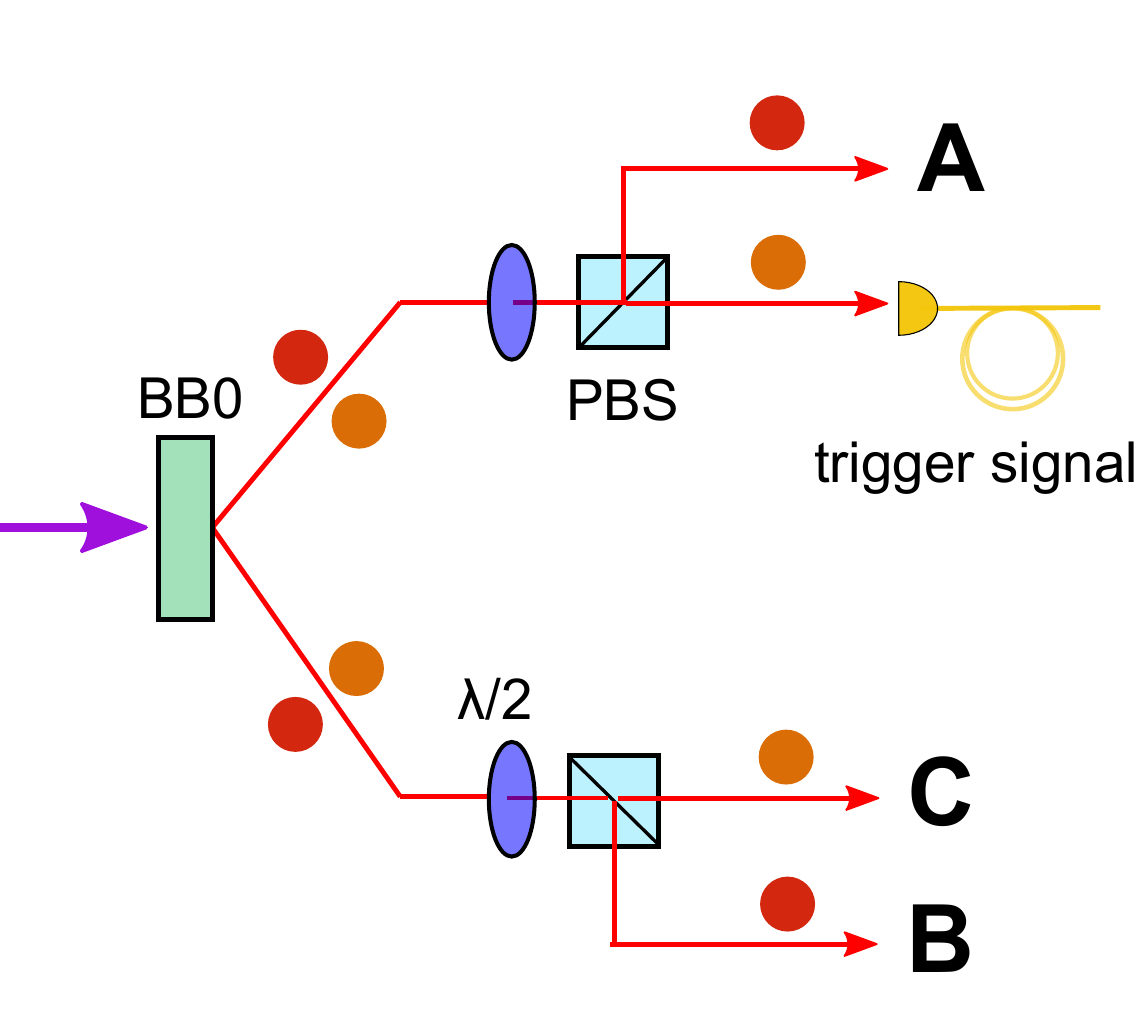}
    \caption{\textbf{Scheme of the three-photon source.} In the experiment, we exploit two-photon pairs emission through spontaneous parametric down-conversion by a nonlinear crystal. Two photons in the pump beam (in purple) are annihilated. Each of the two photons generate a pair, marked with red and orange respectively. The photons are then split into 4 auxiliary arms to post-select the input state of the experiment and perform the preliminary characterization of pairwise indistinguishability.}
    \label{fig:si_source}
\end{figure}

In this section, we provide some further details regarding the three-photon state produced by the source of the experiment. A pulsed laser at $\lambda=392.5$ nm with mean power of $580$ mW, generates two-photon pairs at $\lambda = 785$ nm through type-II spontaneous parametric down-conversion process (SPDC). Each pair is composed of two orthogonal polarized photons, propagating in two different directions (disks of the same color in Fig.\ref{fig:si_source}). 
The two pairs are then split into four modes by two polarizing beam-splitter (PBS). This operation allows us to post-select the state for the experiment, in which we have three photons distributed in the modes A, B, and C plus the trigger photon.
Given the above arrangement of the source, we can perform the HOM dip measurement, namely the recording of simultaneous clicks between single-photon detectors at the output of a beam-splitter with respect to the relative time delay between two photons. This preliminary operation enables us to quantify the actual indistinguishability among photon pairs and to synchronize them for the state manipulation needed by the experiment.

In table \ref{tab:dips_3ph} we report the parameters of the HOM dips retrieved from the measurements. The dip (A, B) is between pairs generated by the same pump photon. Then the HOM test can be performed by recording two-fold coincidence at the outputs of the beam-splitter in which A and B interfere. It is worth noting that the other pairs (A, C) and (B, C) are made up of photons belonging to different pairs, and dip measurements need four-fold coincidence recording, the two beam-splitter outputs plus the two remaining photons acting as heralding signals. The data recorded, two and four-fold coincidences for different relative position of delay lines $\Delta x_{ij}$, have been fitted with the function $C(\Delta x_{ij})=A_{ij}(1-V_{ij}\, e^{-\sigma_{ij}\Delta x_{ij}^2})$, where $A_{ij}$ is a constant, $V_{ij}$ the dip visibility, and $\sigma_{ij}$, that takes into account the dip's width. The $V_{ij}$s for (A, C) and (B, C) are smaller than $V_{AB}$. The reason is that the first two pairs involve interference between photons from different pairs, and the spatial and spectral correlations existing in the multiphoton wavefunction bound the maximum degree of indistinguishability. This preliminary study furnishes the parameters for the surface calculation reported in Fig.3a and Fig.5a in the main text. Furthermore, there values are used to calculate the expected three-photon distributions used to retrieve the overlaps, as we discuss in the next section.
\begin{table}[h]
    \centering
    \begin{tabular}{c c c}
    \toprule
    $(i,j)$ & $V_{ij}$ & $\sigma_{ij}$ \\
    \midrule
      $(A,B)$   & $0.944\pm0.003$ & $(8.7\pm 0.3)\cdot 10^{-5}\, \mu\mathrm{m}^{-2}$ \\
       $(B,C)$   & $0.835\pm0.007$& $(7.0\pm 0.4)\cdot 10^{-5}\, \mu\mathrm{m}^{-2}$ \\
       $(A,C)$   & $0.80\pm0.01$& $(6.2\pm 0.5)\cdot 10^{-5}\, \mu\mathrm{m}^{-2}$ \\
      \bottomrule \\  
    \end{tabular}
    \caption{\textbf{Parameters of the three HOM dips}. In this table we report the parameters of the fit for the three HOM dips performed in the setup of Fig. 2e, in the main text. The relative delays are here expressed in terms of the displacement of the delay lines $\Delta x = c \Delta t$.}
    \label{tab:dips_3ph}
\end{table}

\section{Polarization control of the pairwise overlaps}
\label{app:polarization}
As reported in the main text, the overlap dependence from the polarization degree of freedom is given by the term $\lvert \langle e_i| e_j \rangle \lvert^2 $ 
where $|e_i\rangle$ is the polarization state at the inputs of the beam-splitter that performs the HOM test. 
The control over polarization to prepare states outside polyhedron $C$ is performed by an half-wave plate ($\lambda/2$) placed before the beam-splitter operations.
%In order to control the value of this term, we need to put an half waveplate on one of the beamsplitter input. 
Then the overlap dependence from polarization becomes $\lvert\langle e_i|U_{\lambda/2}|e_j \rangle \lvert^2 $, where $U_{\lambda/2}$ is the transformation operated by the half wave-plate. Since in the experimental apparatus we use in-fiber beam-splitters (FBS) to perform the HOM test, we have to take into account the effects of the input fibers to the polarization of the single-photon states. 
%and it make no possible putting the half waveplate at the input of the beamsplitter. For this reason, we put the plate at the input of the fiber of the FBS.
%In the following, we demonstrate that this two configuration are equivalent and they allow us to control the polarization overlap in the same way. In the configuration adopted in the experimental apparatus, the half waveplate acts on one of the input of FBS. 
Indeed the fiber performs a further transformation that changes the polarization. Hence, the overlap expression becomes $\lvert\langle e_i|U_i^{\dagger}U_j U_{\lambda/2}|e_j \rangle \lvert^2 $, where $U_i$ and $U_j$ are the two different unitary transformations performed by the input fiber of the FBS. By properly compensating the fiber's action via polarization controllers in the apparatus in such way that $U_i=e^{i\phi}U_j$, we can rewrite the overlap expression as $|\langle e_i|U_i^{\dagger}U_i U_{\lambda/2}|e_j \rangle \lvert^2=\lvert\langle e_i|U_{\lambda/2}|e_j \rangle \lvert^2 $. Finally, we retrieve the actual overlap in polarization between the two single-photon states at the input of the FBS.

\section{Three-photon distributions}
\label{app:distributions}
\begin{figure*}[t]
    \centering
    \includegraphics[width=\textwidth]{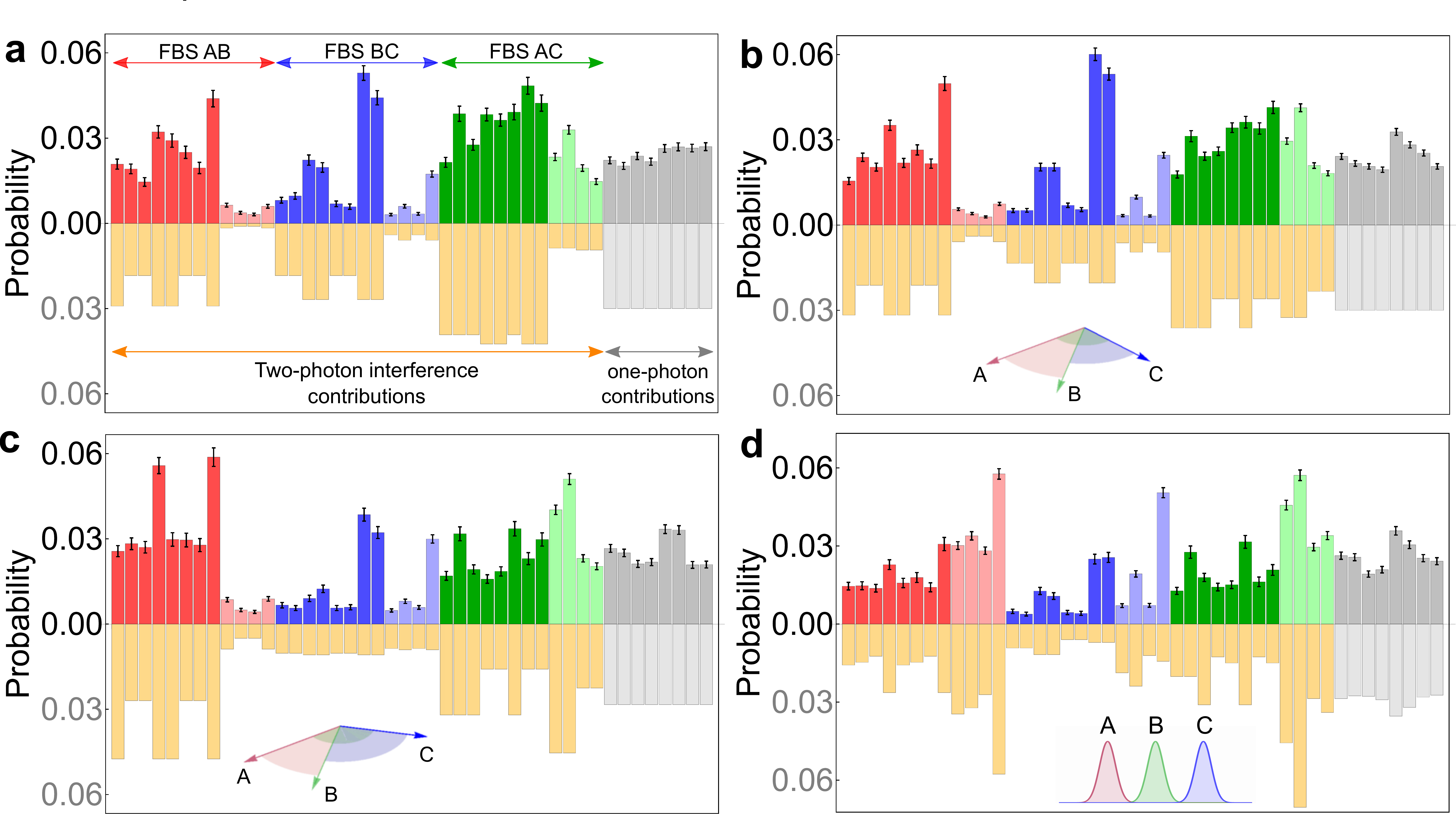}
    \caption{\textbf{Probability distributions of outputs for different preparations of polarization and time delays.} Here we report the three-photon distributions at the output of the device. Above the experimental data, where brighter colors indicate bunched outcomes. Outcomes associated with photon pairs $\{(A,B),\, (B,C),\, (A,C)\}$ are marked respectively with red, blue and green. The grey part of the distribution is insensitive to two-photon interference and so it is irrelevant for the estimation of overlaps. Below the expected distribution according to our model based on preliminary measurements of Hong-Ou-Mandel dips. For each distribution we calculate the total variation distance (TVD) with respect to the theory. Error bars are computed propagating, via Monte Carlo methods, the Poissonian uncertainties associated to photon counts. For each preparation we collected $N\sim 10^4$ four-fold coincidences. States with zero relative delays and different polarization preparation: \textbf{a)} zero relative angles betweeen qubit states, $S_{4}$ in the main text (TVD$=0.122\pm0.005$); \textbf{b-c)} sets of polarization states, $S_2$ and $S_3$ of the main text (TVD$=0.152\pm0.004$, TVD$=0.120\pm0.004$ respectively); \textbf{d)} distribution for single-photon states with the same polarization state but different time delays for violating the dimension witness (TVD$=0.126\pm0.005$)}
    \label{sm_fig:dist}
\end{figure*}
The three pairwise overlaps $\{r_{AB}, r_{BC}, r_{AC}\}$ are obtained from the three-photon distributions. More specifically, they are retrieved from the recorded four-fold coincidences, i.e. three photons at the output of the interferometer of Fig.2e and the trigger photon. In Fig. \ref{sm_fig:dist} we report the distributions for the states measured in the experiment. The overlaps are inferred from the bunching probabilities, the probability to find two photons in the same output of the in-fiber beam-splitter (FBS), for each pair, highlighted by the colors red for (A, B), blue for (B, C) and green for (A, C). These quantities are estimated by additional beam-splitter connected to each output of the interferometer, that perform an approximated photon-number resolving detector. Each distribution for different state preparations in polarization and time delays has been compared with the expected one, calculated according to the actual degree of indistinguishability of the source (see previous section). Furthermore, we have taken into account also losses at propagation and detection stages to obtain a more accurate model of the experimental data. The agreement between the experimental distribution $p^{exp}$ and the theory $p^{th}$ has been quantified via the \emph{total variation distance} (TVD), defined as TVD$=\frac{1}{2}\sum_i |p^{exp}_i-p^{th}_i|$.

\bibliography{bibliography}

\end{document}